# Vector difference equations, substochastic matrices, and design of multi-networks to reduce the spread of epidemics

Harold M Hastings[1] and Tai Young-Taft[2]

[1] Bard College at Simon's Rock, 84 Alford Road, Great Barrington MA 01230, USA
(E-mail: hhastings@simons-rock.edu)
[2] Bard College at Simon's Rock, 84 Alford Road, Great Barrington MA 01230, USA
(E-mail: tyoungtaft@simons-rock.edu)

**Abstract.** Cities have long served as nucleating centers for human development and advancement, c.f. [24]. Cities have facilitated the spread of both human creativity and human disease, and at the same time, efforts to minimize the spread of disease have influenced the design of cities, c.f. [26]. The purpose of this paper is to explore the dynamics of epidemics on networks in order to help design a multi-network city of the future aimed at minimizing the spread of epidemics. In order to do this, we start with the SIR model (susceptible, infected, removed) on a network in which nodes represent cities or regions and edges are weighted by flows between regions. Since the goal is to stabilize the zero infections steady state, we linearize the discrete-time SIR model yielding difference equations for the dynamics of infections at each node and then include flows of infections from other nodes. This yields a vector difference equation $\boldsymbol{v}_{new} = M\boldsymbol{v}$ for the spread of infections. We then generalize the concept of stochastic matrix in order to quantify the dynamics of this update equation. The entries of the update matrix $M$ may vary in time, even discontinuously as flows between nodes are turned on and off. This may yield useful design constraints for a multi-network composed of weak and strong interactions between pairs of nodes representing interactions within and among cities.

## 1 Introduction

Cities have long served as nucleating centers for human development and advancement, c.f. [3-5, 21, 24, 28] and references therein. As Bettencourt *et al.* state: "Cities have long been known to be society's predominant engine of innovation and wealth creation, yet they are also its main source of crime, pollution, and disease. … Many diverse properties of cities from patent production and personal income to electrical cable length are shown to be power law functions of population size with scaling exponents, β, that fall into distinct universality classes. Quantities reflecting wealth creation and innovation have $\beta \approx 1.2 > 1$ (increasing returns), whereas those accounting for infrastructure display $\beta \approx 0.8 < 1$ (economies of scale)." [5]

Schläpfer *et al*. "show that both the total number of contacts and the total communication activity grow superlinearly with city population size, according to well-defined scaling relations and resulting from a multiplicative increase that affects most citizens. Perhaps surprisingly, however, the probability that an individual's contacts are also connected with each other remains largely unaffected. These empirical results predict a systematic and scale-invariant acceleration of interaction-based spreading phenomena as cities get bigger … a microscopic basis towards understanding the superlinear increase of different socioeconomic quantities with city size, that applies to almost all urban systems and includes, for instance, the creation of new inventions or the prevalence of certain contagious diseases." (emphasis added) [21]

Although Arcaute *et al*. [3] question these scaling rules, citing both the role of density and notable exceptions, e.g., the enormous intellectual productivity of Cambridge UK, they also support the central role of cities.

Finally, the there is a two-way relationship between cities and disease: as Stinson [26] observed, "Health and Disease Have Always Shaped Our Cities." For example, Olmstead and Vaud's development of Central Park (an 843 acre/ 341 hectare park) in New York City was in part driven by cholera epidemics in the early to mid-1800's.

Stinson then asked "What Will Be the Impact of COVID-19?", a question which motivated the thrust of the present paper. Perhaps the city of the future and the inter-urban network of the future will be in part physical, in part virtual, thus a milti-network.

We start with the SIR model (susceptible, infected, removed) (c.f. [23] for a review on a network. The SIR model has been widely used to stud the spread of Covid-19 (c.f. [9] and references therein). Biswas was *et al.* explored the early spread of Covid-19 in China with the SIR model on a network [6], obtaining a power law fit to the contacts. Here we develop a simple matrix/network extension of the SIR model near the desired zero infections steady state.

The SIR model considers three states: susceptible, infected, removed. We shall not distinguish here between "infected" and "infectious," since this distinction does not affect dynamics near the desired zero infections steady state. In addition, we simplify the model by assuming that infected individuals either gain long-term immunity or pass away. Under our simplifying assumptions we have a simple flow diagram (Fig. 1, below) and differential equation for the number infected (eqn. 1, below). (If immunity were only temporary, some of the immune population would become susceptible, and we would add a flow from the removed compartment to the susceptible compartment.)

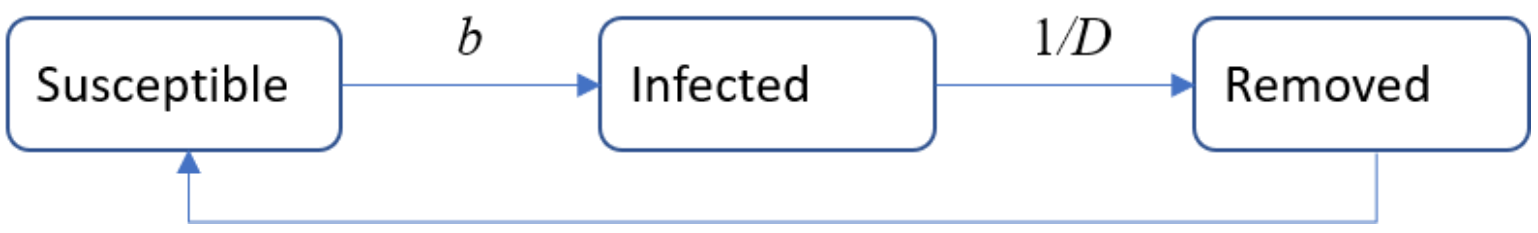

**Fig. 1. Flow diagram for the SIR model**. Arrows represent flows from susceptible infected (as susceptible individuals become infected), from infected to removed (as infected individuals become no longer infected, some of whom pass away and others recover), and from removed to susceptible (recovered individuals who lose or never had immunity). Since Covid-19 appears to show long-term immunity in most recovered individuals [11, 22], we may ignore flows from removed to susceptible. Flows into the infected compartment depend upon interactions between susceptible and infected individuals, parametrized by an effective transmission rate *b*; flows out of the infected department scale inversely with and the duration *D* of infection. See text, eqn. (1).

Flows into and out of the infected state (compartment in engineering language) are given by the standard SIR equation

$$dI/dt = bIS - I/D \qquad (1)$$

Here *I* denotes the number infected, *S* the number susceptible, *b* an effective transmission rate (described below) and *D* the duration of infection. Our goal here to make the steady state with zero infected ($I = 0$) Lyapunov stable equilibrium as in May and Anderson [20]. As they observed, "simple mathematical models of the transmission dynamics of HIV help to clarify some of the essential relations between epidemiological factors." We shall rewrite the flow from susceptible to infected in terms of probability *p* that a contact yields an infection, and the effective contact rate *c* between susceptible individuals and the probability *s* that a contact is susceptible, that is, $bS = pcs$. As illustrated in Fig. 2 below, the SIR model for infections then becomes

$$\begin{aligned} dI/dt &= pcsI - I/D \\ &= (pcs - 1/D)I \qquad (2). \end{aligned}$$

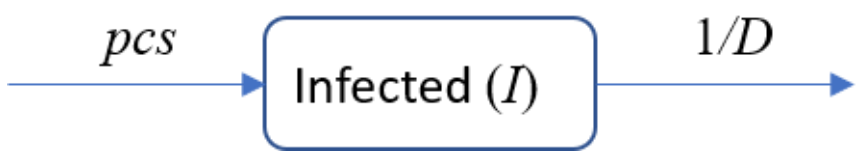

**Fig. 2. Graphical representation of the SIR model (eqn. (2))**. Flows are *pcsI* (into the infected compartment) and (1/*D*)*I* (out of the infected compartment) for a net flow of (*pcs* – 1/*D*)*I*.

We shall now explore the role of controls (masking, distancing, travel limits, shutdowns, etc.) aimed at controlling the cascading spread of Covid-19 through the use of a robust, simple model, namely a matrix/network extension of the above SIR model.

## 2 Linearizing the SIR model

In order to develop the matrix model, we first linearize the SIR model (eqn. (2)) by assuming that the probability $s$ that a contact is susceptible is constant. We then discretize the linearized SIR model by the Euler method, obtaining the difference equation

$$I(t+\Delta t) = I(t) + (pcs - 1/D)\, I(t)\, \Delta t$$
$$= (1 + (pcs - 1/D)\Delta t)\, I(t) \qquad (3).$$

**Remarks.**

(1) The parameter $R_t$ (c.f. the basic reproductive rate $R_0$ in [20]) is readily obtained by setting the time step $\Delta t$ in eqn. (3) equal to the duration of infection $D$. The following simple calculation then implies that

$$R_t = pcsD \qquad (4)$$

is the growth rate of infections over the duration of infection $D$ as the natural time scale:

$$I(t+D) = (1+( pcsD-1)\, I(t)$$
$$= pcsD\, I(t) \qquad (5).$$

(2) Eqn. (6) yields a condition for herd immunity, namely, that the growth rate

$$R_t = pcsD < 1 \qquad (6),$$

in which case the number of infected, $I$, decreases at least exponentially as

$$(\mathrm{pcsD})^{t/D} \qquad (7).$$

The required limit in growth rate $R_t$ (eqn. (6)) can be can be achieved in several ways or by an appropriate combination thereof:

a. Reducing the probability $p$ that a contact results in an infection through masking and/or physical (misnamed social) distancing. Of course, relaxing masking and/or physical distancing will increase $p$, potentially causing the growth rate $R_t$ to increase above 1, ending (apparently temporary) herd immunity.

b. Reducing the effective contact rate $c$ between susceptible and infected individuals, for example by making some contacts virtual and/or by otherwise reducing the incidence of "superspreading"/ "spreading" events in a multi-networked city of the future. May and Anderson [20] emphasized the role of the distribution of contact rates in determining the effective contact rate (for the spread of HIV in a male homosexual population):

$$c = \mu + \sigma^2/\mu \quad (8),$$

the mean + the ratio of the variance to the mean, which we rewrite perhaps more simply as

$$c = \mu\,(1 + (\sigma/\mu)^2) \quad (9).$$

The effective contact rate is thus increased by the square of the coefficient of variation, similarly emphasizing the role of variability in contact rates. Of course, eqns. (8) and (9) "blow up" if the contact rate has a fat tailed distribution with infinite variance.

Similarly, reducing contacts through travel restrictions (the effects of travel modeled by off-diagonal elements in a matrix update equation (eqns. (10ff), below).

c. Reducing the proportion of susceptibles $s$ by vaccination.

## 3 Substochastic matrices, products with non-negative vectors and a matrix/network model.

In this section we extend the linearized, discrete-time SIR model (eqn. (4)) to networks.

**Notation.** Since the symbol $i$ is typically used for indexing vectors and matrices, we write $\boldsymbol{v}$ for the vector whose entries denote the number of infected individuals in nodes (cities, regions). We shall also follow standard notation $I$ the identity matrix from here on, and $M$ for the matrix of rate coefficients in our matrix/ network model. Thus diagonal entries are given by rate coefficients in the SIR model (eqn. (4)). Off-diagonal entries in $M$ represent rates at which infections in one node/city/region generate infections in another node/city/region; see Fig. 2, below.

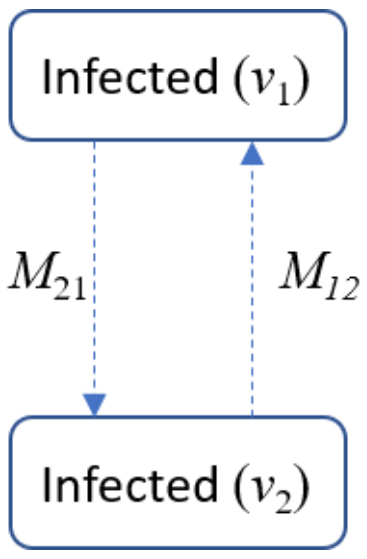

**Fig. 2. Infections generated by connections in the matrix/network model.** This small, "toy" model consists of two nodes, representing cities or regions. In this representation $M_{21}$ represents the rate at which one infection in node 1 generates new infections at node 2, that is $M_{21}\Delta t$ new infections at node 2 after one time step $\Delta t$. Analogously $M_{12}$ represents the rate at which each infection at node 2 generates new infections at node 1.

We obtain an update rule $\boldsymbol{v}(t+\Delta t) = \boldsymbol{v}(t) + M\boldsymbol{v}(t)$, or more simply

$$\boldsymbol{v}_{\text{new}} = (I+M\Delta t)\boldsymbol{v} \qquad (10),$$

Here the diagonal elements represent local (within node/city/region) SIR interactions

$$M_{ii} = p_i c_i s_i - 1/D \qquad (11),$$

and the off-diagonal elements $M_{ij}$ represent the rate at which infections at node $i$ generate additional infections at node $j$. Fig. 3, below, is a graphical representation of the matrix/network SIR model.

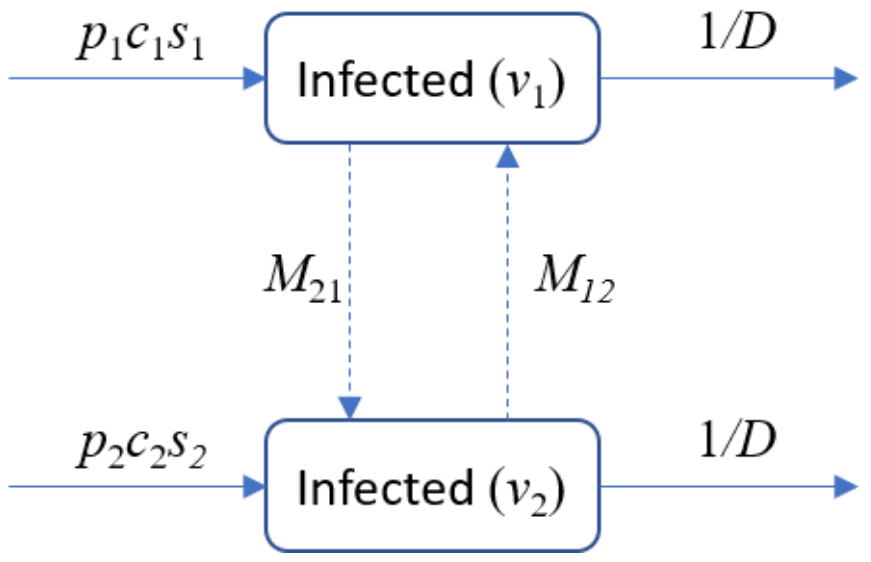

**Fig. 3. Graphical representation of the matrix/network SIR model.** This small, "toy" model consists of two nodes, representing cities or regions. As in Fig. 2, $M_{21}$ represents the rate at which one infection in node 1 generates new infections at node 2, that is $M_{21}\Delta t$ new infections at node 2 after one time step $\Delta t$, analogously for $M_{12}$. In addition, the rate of change $\boldsymbol{v}_1$, the number infected at node 1 is simply $(p_1c_1s_1-1/D)\boldsymbol{v}_1$ as in eqn. (11), analogously for node 2; see also Fig. 1, above.

**Interpretation and consequences.** Suppose $\boldsymbol{v}$ is the unit vector with a 1 in the $j^{\text{th}}$ row, representing one infection in node $j$ of the network. Then $M\boldsymbol{v}$ is simply the $j^{th}$ column of $M$: $(M\boldsymbol{v})_i = M_{ij}$. The $j^{th}$ column sum $\Sigma_i M_{ij}$ is the "rate constant" at which infections at node $j$ drive a change in the number of infections throughout the entire network. As a consequence, if all column sums of the update matrix $M$ are negative, then the total number of infections in the network will decrease after a time step $\Delta t$. This observation suggests that the column sums play a role in quantifying the growth or decay of infection in a matrix/network model analogous to the role of $R_t$ in quantifying local dynamics: (column sums < 0) ~ ($R_t < 1$).

The analogy (column sums < 0) ~ ($R_t < 1$) can be simplified by rewriting the update equation $\boldsymbol{v}_{\text{new}} = (I+M\Delta t)\boldsymbol{v}$ (eqn. (10)) as

$$\boldsymbol{v}_{\text{new}} = (I+M\Delta t)\boldsymbol{v} = A\boldsymbol{v} \qquad (12),$$

where

$$A = I+M\Delta t \qquad (13).$$

In the case of a trivial network (one node, no connections), and time step $\Delta t = D$, the duration of infection, $A$ becomes a scalar and $A = R_t$ (eqn. (4), above). The stability criterion $R_t < 1$ then suggests the following definition (generalization).

**Definition 1.** A non-negative matrix is column-substochastic if its maximum column sum is less than 1.

Thus, whenever the update matrix $M$ in the update rule is column-stochastic, one infection yields < 1 infections in total at the next time step, irrespective of location.

**Note.** The $\ell^1$ norm of a non-negative vector is simply the sum of its entries. Here the $\ell^1$ norm of $\boldsymbol{v}$ is simply the total number of infections.

We need a relatively straight-forward theorem on left multiplication by a column-substochastic matrix. We give a short proof (a straight-forward calculation) here since we did not find a proof in the literature.

**Theorem 1.** Left multiplication by a column-substochastic matrix with maximum column sum $\leq k$ reduces the $\ell^1$ norm of a non-negative vector by a factor $\leq k$.

**Proof.** Let $\boldsymbol{v}$ be a non-negative vector and $A$ a column-substochastic matrix. We calculate the $\ell^1$ norm of $A\boldsymbol{v}$, namely $\Sigma_i (A\boldsymbol{v})_i$ as follows.

$$\Sigma_i (A\boldsymbol{v})_i = \Sigma_i (\Sigma_j A_{ij}\, \boldsymbol{v}_j)$$
$$= \Sigma_{ij} A_{ij}\, \boldsymbol{v}_j$$

$= \Sigma_j (\Sigma_i A_{ij}) \boldsymbol{v}_j$

$\leq \Sigma_j k \boldsymbol{v}_j$

$= k\Sigma_j \boldsymbol{v}_j$ (14),

namely $k$ times the $\ell^1$ norm of the non-negative vector $\boldsymbol{v}$. **Done.**

**Remarks.** The Gerschgorin circle theorem yields an analogous bound for the eigenvalues of a non-negative matrix, based on row sums. However, in the spirit of Cohen and Newman's results for products of random matrices [7, 8], Theorem 1 has the advantage that it can be generalized to products of column-stochastic matrices, which themselves may vary in time.

**Corollary 1.** Repeated application of the update rule $\boldsymbol{v}_{\text{new}} = A\boldsymbol{v}$ (where $A = I+M$) to a non-negative vector $\boldsymbol{v}$, represented by the product of a sequence of column-substochastic matrices $\{A(i)\}$, which may vary in time, and all with maximum row sum $\leq$ k < 1, namely

$$\boldsymbol{v}(t+n\Delta t) = A(n)\, A(n-1) \ldots A(2)\, A(1)\boldsymbol{v}(t) \quad (15),$$

reduces the $\ell^1$ norm of $\boldsymbol{v}$ by a factor of at most $k^n$.

**Corollary 2.** Thus, near the zero-infections steady state, absent introduction of infections from outside the network, the total number of infections decays at least as fast as $k^n = k^{t/\Delta t}$ (exponentially), that is, with time scale

$$\tau \leq -\Delta t \ln k \quad (16).$$

.More generally we have the following.

**Corollary 3.** Repeated application of the update rule $\boldsymbol{v}_{\text{new}} = A\boldsymbol{v}$ ($A = I+M$) to a non-negative vector $\boldsymbol{v}$, represented by the product of a sequence of column-substochastic matrices which may vary in time, as in eqn. (15) above, reduces the $\ell^1$ norm of $\boldsymbol{v}$ by a factor of at most $k_{\text{g.m.}}{}^n$ where $k_{\text{g.m.}}$ is the geometric mean of the corresponding maximum column sums.

**Corollary 4.** Thus, near the zero-infections steady state, provided that

$$k_{\text{g.m.}} < 1 \quad (17),$$

and absent introduction of infections from outside the network, the total number of infections decays exponentially at least as fast as $k_{\text{g.m.}}{}^n = k_{\text{g.m.}}{}^{t/\Delta t}$, that is, with time scale

$$\tau \leq -\Delta t \ln k_{\text{g.m.}} \quad (18).$$

Of course, Theorem 1 its corollaries approximate converses, based upon the following definition of column-superstochastic matrices. Since details are similar, here we simply state the converse and resulting time scales for growth without proof.

**Definition 2.** A non-negative matrix is column-superstochastic if its minimum column sum is greater than 1.

**Theorem 2.** Left multiplication by a column-superstochastic matrix with minimum column sum $\geq K$ increases the $\ell^1$ norm of a non-negative vector by a factor $\geq K$.

Thus, repeated application of the update rule $\boldsymbol{v}_{\text{new}} = A\boldsymbol{v}$ to a non-negative vector $\boldsymbol{v}$, represented by the product of a sequence of column-substochastic matrices which may vary in time, and all with minimum row sum $\geq$ K > 1, as in eqn. (15) above, causes the number of infections to grow at least as fast as $K^n$ (exponentially), that is with time scale

$$\tau \leq \Delta t \ln K \qquad (19).$$

If we relax the criterion and consider only the geometric mean $K_{\text{g.m.}}$ of the corresponding maximum column sums, the number of infections grows at least as fast as $K_{\text{g.m.}}{}^n$ (exponentially), that is with time scale

$$\tau \leq \Delta t \ln K_{\text{g.m.}} \qquad (20).$$

## 4 Discussion

In sections 2 and 3 we extended a linearized SIR model to a corresponding matrix network model, and extended the usual criterion $R_t < 1$ for infections to decline, ultimately stabilizing the zero infection steady state to a corresponding criterion for in the matrix/network model (Corollaries 1 and 2). Here we discuss that criterion and its consequences.

Recall the update rule (eqn. 12)

$$\boldsymbol{v}_{\text{new}} = (I+M\Delta t)\boldsymbol{v} = A\boldsymbol{v},$$

where $\boldsymbol{v}$ for the vector whose entries denote the number of infected individuals in nodes (cities, regions) in the network, the diagonal entries

$$M_{ii} = p_i c_i s_i - 1/D$$

represents SIR dynamics at node $i$, and the off-diagonal entries parametrize flows between regions. We developed a variety of stability criteria for driving the infections to 0, of which Corollary 1 is the simplest: the update matrices $A = I+M\Delta t$ are uniformly substochastic, with all column sums of all update matrices $A$ uniformly bounded by a constant $k < 1$. This criterion combines the generation of new infections at each node with infection "exported" to other nodes: the following analog of $R_t$, namely 1+ the total number of infections generated throughout the network by a single infection at any node, less the recovery rate $1/D$, given by the corresponding column-sum of $M$, must be uniformly bounded by a constant $k < 1$. The bound on column sums can be considered as a granular version of a bound on $R_t$.

### 4.1 Consequences for control and network design

We first discussed controlling the spread of coronavirus locally (at one node) by making $R_t = pcsD < 1$ at that node, through the combined effects of (a) Slowing the spread by reducing the probability that a contact results in an infection through masking and/or physical (misnamed social) distancing, (b) Reducing the effective contact rate between susceptible and infected individuals, for example by making some contacts virtual and/or by otherwise reducing the incidence of "superspreading"/ "spreading" events in a multi-networked city of the future, and (c) reducing the proportion of susceptibles by vaccination.

Our matrix/network models introduce more granularity: for stability an increasing rate of infections at a node, namely $p_jc_js_j$, must be balanced by a similar reduction the rate of infections generated by infections at that node, the latter parametrized by the sum of off-diagonal elements in the corresponding column of $M$ (or $A$ since $A= M+I$), for example, through localizing or virtualizing interactions, restrictions on travel, or reduction of the size of events where interactions occur. Similar considerations apply in reducing the rate of growth of infections in cases where infections are growing (column sums of update matrices $A > 1$, analogs of $R_t >1$) to reduce the strain on resources by "flattening the curve."

Moreover, the spread of Covid-19 seems to have the hallmarks of a cascading failure [10], here initially localized loci of infection which spread more rapidly than the system reacts to contain or mitigate the spread (described for example in [14]). Decentralized (modular) organization has been shown to promote survivability from "cascading failures in power grids." [16] discussed how decentralized (modular) organization and "reciprocal altruism" promote survivability from "cascading failures in power grids." The modular organization of the US (and perhaps any other) power grids, and lessons learned from cascading failure leading to the Northeast US blackout of 2003 [18] may provide ideas for a modular organization of networks in cities of the future, in which case signals provided by matrix models such as those discussed above (e.g., eqn. (9)) might provide fast enough warning signals to contain the spread of future epidemics from hot spots. For example, one might seek to avert the transition from substochastic dynamics (decay of infection) to superstochastic dynamics (growth of infection), both throughout the network, by introducing travel restrictions, reducing off-diagonal elements in the update matrix. See also the discussion of the "smart electric grid of the future" [2].

### 4.2 Limitations, future work and conclusions

In the present paper have discussed only the simplest matrix/network extension of the classic SIR model, and looked only for qualitative correspondence to the data. In the future one should explore extensions to more general models and explorations of fitting data (and explanations for failure to fit the data, e.g., evolving social behavior, c.f. [9].

Moreover, one might explore what might be interesting non-linear extensions to the present discussion. The May-Wigner transition from instability to stability as a function of the strength of interactions in a network with linear interactions [1, 7, 8, 19] has also been observed in nonlinear dynamical systems [12, 17, 25]. Network structure also plays a role in stability, both small-world organization [27] and more complex network structures (c.f. [15]), and should be considered in understanding the role of social behavior and urban design in the spread of epidemics.

We also plan to extend our matrix/network model to a stochastic model to study the propagation and growth of random infections in a matrix/network model (c.f. [13]), and to develop a criterion for "circuit breakers" to limit cascading spread (see the discussion of the power grid, above). Both projects project may help in "flattening the curve" to help avoid overrunning the limits of medical resources.

### 4.3 Summary

In summary, we have developed a model for the spread of epidemics which over networks, which includes both local (within node/city/region) SIR dynamics and flows of infections between nodes. We have also provided a natural criterion for the zero infections steady state to be stable: expressed mathematically that all column sums are less than 1 (substochastic), and interpreted that one infection anywhere in the network yields less than one new infection (on average) over the entire network. This approach may be a small step towards models to help design "smart cities" of the future with an idea of preventing, containing, or at least mitigating future pandemics.

*Acknowledgement.* We thank Jenny Magnes (Vassar College) for a helpful comments and edits on a prior version of this manuscript.